\documentclass[12pt]{iopart}
\usepackage{epsf} 
\usepackage{graphics}
\usepackage{amssymb}
\usepackage{iopams}

\begin{document}

\title[ENHANCED GREENHOUSE EFFECT]{NEW EVIDENCE OF AN ENHANCED GREENHOUSE EFFECT}

\author{Rasmus E. Benestad}

\address{The Norwegian Meteorological Institute}
\ead{rasmus.benestad@met.no}
\begin{abstract}
The state of earth's climate is constrained by well-known physical principles such as energy balance and the conservation of energy. Increased greenhouse gas concentrations affect the atmospheric optical depth, and physical consistency implies that changes in the energy transfer in terms of infra-red light must be compensated by other means of energy flow. Here, a simple heuristic and comprehensive model is used to interpret new aspects of real-world data. It is shown that trends in tropospheric overturning activity and the estimated altitude where earth's bulk heat loss should place are two independent indicators of climate change. There has been increased vertical overturning in the middle and upper parts of the troposphere since 1995 on a global scale. Greater overturning compensates for reduced radiative energy transfer associated with increased optical depth. An increased optical depth is also expected to raise the altitude from where planetary bulk heat loss takes place according to the heuristic model, and an estimated trend of 40m/decade is consistent with a surface warming rate of 0.2K/decade. The simple comprehensive model of the greenhouse effect can also account for feedback processes, changes in the hydrological cycle, and may assist the wider scientific community in understanding of the main principles involved. 
\end{abstract}


\bibliographystyle{Science} 

\maketitle

\section{Motivation}

A number scientists still doesn't acknowledge the link between increased levels of GHGs and climate change \cite{Anderegg2010}, having not been persuaded by the scientific literature \cite{Oreskes2004,IPCC-AR4} supporting the notion of an anthropogenic global warming (AGW). The essential principles of the greenhouse effect (GHE) have been long understood \cite{Weart2003} in terms simple reductionistic models \cite{Fleagle80,Houghton91,Peixoto92,Hartmann94,Lacis2010,Pierrehumbert2011}, however, these models don't provide a comprehensive description of the complexity associated with an AGW. State-of-the-art global climate models (GCMs) again are too complex for comprehension. 

\subsection{A simplified physical picture}

One key concept of the GHE that can give us further insight, however, is that the energy flows through the universe, being generated inside the sun and intercepted by the planets \cite{Trenberth2004}. When the energy leaves the terrestrial system, it must be in the form of electromagnetic radiation, since space is a vacuum and does not provide a medium for other type of energy transfer \cite{Pierrehumbert2011}. 

The radiative energy emitted from any object is described by Planck's law, relating electromagnetic energy loss to the emission temperature ($T_e$). The rate of heat loss must also equal the rate of energy received from the sun for a planet in energetic equilibrium. The planetary energy balance can then be described by the simple equation 

\begin{equation}
S_0 (1-A)/4 = \sigma T_e^4,
\label{eq:1}
\end{equation}

\noindent where $\sigma=5.67 \times 10^{-8}W/(m^2 K^4)$ is the Stefan-Boltzman constant, $A \sim 0.3$ is the albedo, and $S_0=1366W m^{-2}$ is the 'solar constant'. The left hand side of this equation represents the energy input while the right hand side describes the heat loss. Equation \ref{eq:1} can be compared with observed surface temperatures in our solar system to demonstrate predictive skill\footnote{See Figure S1 in the supplementary material}.

Most of the energy from the sun is absorbed at the planet's surface, as earth's atmosphere is transparent to visible light. Equation \ref{eq:1}, however, gives a value of 254K for the emission temperature whereas earth's observed global mean surface temperature is 288K. The atmospheric radiation transfer is rather complicated, as the gases absorb light at different wavelength according to detailed features in their line spectra. Part of the infrared light escaping the earth's atmosphere originates from vertical levels near the surface and parts are emitted from heights near the tropopause \cite{Pierrehumbert2011}. 

A simplified picture of the situation is to treat all the infrared light (IR) as one bulk heat loss, as in equation \ref{eq:1}, and defining the bulk emission level $Z_e$ as the region where the temperature equals the emission temperature $T=T_e$ \cite{Houghton91}. A similar treatment of the bulk emission and its vertical level is also the basis for the discussion on the 'saturation effect' in the report 'The Copenhagen Diagnosis' (2009)\cite{TheCopenhagenDiagnosis2009}.

In this simplified picture, the bulk of the heat loss to space cannot take place at the ground level where $T=288K$ because this would violate the energy balance (equation \ref{eq:1}). The temperature drops with height and equals the emission temperature of 254K at around 6.5 km above the ground\footnote{See the Figure S2 in the supplementary material}, and this is where earth's bulk heat loss must take place according to equation \ref{eq:1}. This altitude is henceforth referred to as the 'bulk emission level' $Z_{T254K}$. The requirement of energy conservation everywhere implies that there has to be a constant energy flow from the ground, where the sun's energy is deposited, to $Z_{T254K}$, where it escapes to space. There are different ways through which the energy may flow, such as convection, dynamical adjustments (gravity waves), and radiative energy transfer. 

The GHE consists of the air being opaque to light in the long-wave range while transparent for short-wave radiation. The transparency of a medium is described by Beer's law, which relates the optical depth to the concentration of the absorbing medium and the medium's absorbing capacity \cite{Fleagle80,Peixoto92}. The IR light is expected to be absorbed and re-emitted multiple times before its energy reaches the emission level where it is free to escape to space \cite{Pierrehumbert2011}. Hence, for an observer viewing the earth from above, the altitude of earth's bulk IR light source would be located at increasing heights $Z_{T254K}$ with greater greenhouse gas concentrations (GHGs), as the depth to which the observer can see into the atmosphere gets shallower for more opaque air. 

Moreover, the GHE affects the energy flow between the surface and the emission level, and a deeper optical depth, due to increased absorption, is expected to act as resistance for the radiative energy transfer, everything else being constant. Reduced radiative energy flux must be compensated though increased temperatures, latent/sensible heat fluxes, or more energetic gravity waves. 

\section{Methods \& Data}

Convection is characterised by a circulation pattern of rising and sinking air masses. The atmospheric vertical volume transport takes place through cells of updraught and subsidence, however, these may not be coherent or stable in time and space. It is therefore important to find a metric that doesn't assume patterns that are stable in time. 

Here the global spatial variance in the vertical volume transport for a given moment in time is used to represent the state of vertical overturning on a global scale. Sensible and latent heat transfer are related to the vertical motion, but not completely determined by the flow, as the atmospheric vapour content and temperature gradients also are important factors. But water vapour content is poorly constrained, due to uncertainties associated with clouds and precipitation. 

The vertical velocity ($v_z$), on the other hand, is related to horizontal air flow through divergence, convergence, and variation in well-described quantities such the geopotential heights. Hence the monthly mean $v_z$ from 21 model levels between 1000hPa and 250hPa (which corresponds to $\sim$0m and $\sim$12500m above sea level) from the European Centre for Medium-range Weather Forecasts (ECMWF) interim reanalysis (ERAINT \cite{ERAINT})  was used provide a metric for the global atmospheric overturning (Jan-1989-Oct-2010)\footnote{See the supporting material for more details and listings of scripts used for the analysis}. 

The reanalysis could only represents large-scale flow that were explicitly resolved by the model (horizontal resolution is $1^\circ \times 1^\circ$), and the overturning metric ($var(a \times v_z)$) for a given date was estimated for each vertical level by multiplying the ERAINT grid-box area (a) with the grid-box value $v_z$, and then take the variance over all contemporary grid-box values. Hence $var(a \times v_z)$ describes deviations in the vertical volume flow without assuming stationary or coherent structures.

\section{Results}

Figure 1 shows $var(a \times v_z)$ for 3 different levels in the troposphere. The variability above $\sim$1000 m a.s.l has increased since 1995, with most pronounced increase in the middle troposphere ($\sim$1km--6.5km a.s.l.; black). The upward trend in the middle atmosphere is consistent with the notion that increased convection compensates for reduced radiative transfer between the ground and $Z_{T254K}$. The upper layer above $\sim$6.5km a.s.l. also exhibits a smaller trend, but this layer may be affected by the convective activity below. Furthermore, part of the IR light is also emitted from levels higher up than 6.5km \cite{Pierrehumbert2011}, although the upper level is associated with drier air and hence lower latent heat transport. 

The global variations in $var(a \times v_z)$ in the lower 1000m, however, have an inter-annual time scale, but nevertheless exhibit no clear association with the El Ni\~{n}o Southern Oscillation. This layer embeds most of the planetary boundary layer (PBL), which is affected by the surface friction and turbulent mixing, whereas the air above represents the free atmosphere. In the lower PBL, the main energy transport takes place through small-scale turbulence with short time scales which may not be well represented by monthly mean values from the atmospheric model used for the reanalysis, may not have been assimilated well due to scarce observations for that level, and are subject to incomplete understanding. 
	
The 254K isotherm $Z_{T254K}$ represents the altitude where earth's bulk heat emission takes place, and an upward trend of 40m/decade is consistent with a deeper optical depth (Figure 2; black curve). This means that an observer in space would see IR radiation emanating from a shallower atmospheric depths as the GHGs have increased over time\cite{IPCC-AR4}\footnote{Also see the supplementary material}. 

The blue curve in Figure 2 shows the vertical integral of total column water vapour ($q_{tot}$), exhibiting a weak upward trend, which together with increased overturning supports the notion of an enhanced vertical latent heat transport. The excursions in both $q_{tot}$, and $Z_{T254K}$ in 1991--1992 and 1997--1998 are associated with ENSO, and demonstrate how the optical depth is affected by changes in the atmospheric moisture. 

\section{Discussion}

The trend analysis for the 254K isotherm $Z_{T254K}$ can be considered as an extension of Santer et al. (2003) \cite{Santer2003b}, who reported a robust, zero-order increase in tropopause height over 1979--1997 in two earlier versions of re-analyses, which they interpreted as an integrated response to anthropogenically forced warming of the troposphere and cooling of the stratosphere. Here, this aspect is put into the physics context of energy flow, where the metric $Z_{T254K}$ is interpreted as a mean level for the bulk emission. This type of representation is a simplification on par with 'model physics' such as parameterisation of clouds in the GCMs themselves. Such simple models are strictly not correct but may nevertheless be useful. Hence, the emission level diagnostic is not actually an emission level, it's just the height of the temperature that corresponds to the bulk emission level on the right hand side of  equation \ref{eq:1}. One objective here was to provide a simple picture of an enhanced GHE.

The trend analysis in the atmospheric overturning metric presented here appears to be inconsistent with the previous conclusions \cite{Vecchi2007} suggesting that the strength of the atmospheric overturning circulation decreases as the climate warms. 

On the other hand, if one considers the energy flow from the surface to the emission levels higher aloft, and that an increases optical depth diminished the vertical energy transfer associated with IR radiation, a reduction in convective activity is surprising, as the flow of energy must be continuous and any divergence in the total flux will lead to deficit or surplus of heat. A simulated warming in the upper troposphere \cite[Figures 9.1 \& 9.2]{IPCC-AR4} nevertheless suggests an enhanced emission of IR from greater heights, which must be supported by an increased vertical energy flow. It is, however, possible that an increased latent heat flux, due to increased atmospheric moisture, can maintain this even if the overturning were to slow down. It is also conceivable that energy is transferred through other means of energy transfer, such as the propagation of gravity waves which tend to be represented by the means of parameterisation in atmospheric models. 

A change in $var(a \times v_z)$ can also be linked to reports of changes to the Hadley cell \cite{Seidel2008}, and changes in convective activity may have some impact on the hydrological cycle as these tend to involve cloud formation. Furthermore, persistent high- and low-pressure systems, an aspect of large-scale convection and subsidence, are associated with weather extremes such as drought, heat waves, cold winters, and floods. 

The variations in $q_{tot}$ must have an impact on the hydrological cycle, as excess moisture associated with the excursion must precipitate out. The analysis also indicates that there is enhanced latent heat transfer associated with El Ni\~{n}o, and the increased atmospheric moisture is consistent with a more moist adiabat and a reduced temperature height dependency $|dT/dz|$\footnote{See Figure S2 in the supplementary material}. A reduction in $|dT/dz|$ is consistent with the negative lapse-rate feedback \cite{IPCC-AR4}, but there is also a long-term increase in $q_{tot}$ consistent with a positive feedback of increased optical depth associated with $H_2O$. A 40m/decade increase in $Z_{T254K}$ translates to a surface warming of ~0.2K/decade at the ground, given dT/dz = -5K/km\footnote{See Figure S2 in the supplementary material}. 

Clouds play an important part, as they are a product of convection. They represent one of the big uncertainties in climate modelling, and different GCMs provide different accounts about their roles in terms of the albedo \cite{IPCC-AR4}. A climate change may conceivably introduce changes in cloud distributions and properties, but since they are not well-resolved in the GCMs, only a crude description is available. A description of clouds is also included in the ERAINT re-analyses, and any discrepancy in their representation may affect the hydrological cycle and hence the results reported here.

A trend in the 254K isotherm or atmospheric overturning alone does not rule out explanations such as changes in the solar forcing. However, there is no long-term trend in the solar activity over the last $\sim$50 years \cite{Benestad2005b,Lockwood2007}\footnote{Also see the Figure S4 in the supplementary material}, ruling out a significant solar influence. The sunspot number $R_z$ exhibits no similarity between the level of solar activity and the atmospheric overturning. 

Past studies on long-wave radiation spectra of the earth in 1970 and 1997, on the other hand, support the notion of a strengthening of the GHE \cite{Harries2001}. The outgoing long-wave radiation (OLR) represented by the right-hand side of equation 1 is affected by both $S_0$ and $A$, and OLR measured from the Earth Radiation Budget Experiment (ERBE) over the period 1985 and 1998 suggest a mean value of $254 Wm^{-2}$ and an upward trend of $3.9 Wm^{-2}/decade$ \cite{Wang2002}. However, 40\% of this was attributed to changes in cloud vertical distributions, and a change in cloud effective emissivity of could account for the remainder. A mean OLR of $254 Wm^{-2}$ also suggests an emission temperature of 259K as opposed 254K, but an error of 2\% is remarkably good for the simple energy balance equation used here. 

ERAINT provides the to-date best picture that we have of the atmosphere and an improvement over previous studies \cite{Santer2003b}, but there are some caveats as the introduction of new instruments, such as satellite instruments, will result in inhomogeneity. Furthermore, unresolved processes, such as cumulus convection may not be well-represented, and only monthly means were used for the vertical velocity. Nevertheless, increased atmospheric overturning and emission level altitude support the conceptual picture of an enhanced GHE, for which the energy flow between the ground and the emission level must be constant. 

Normally, trend analysis based on 21 years will be too short for inferring significant and systematic long-term changes. However, if the analysis can be tied to our understanding of the physics and causation, it may nevertheless be able to shed light on the situation. Here, the change in the $Z_{T254K}$ and  $var(a \times v_z)$ over the period January-1989--October-2010 provides a consistent picture in terms of increased optical depth and elevated emission levels.

\section{Conclusions}

Physical constraints provided by the simple conceptual model may give us untraditional finger prints of an enhanced greenhouse effect. This frame work can also embed more sophisticated explanations, incorporate effects from feedback processes through $dT/dz$, response in $Z_{T254K}$, and $A$, and provides a link to the hydrological cycle and extreme weather events associated with convection. Another interesting question is whether these new metrics and their interpretation through simple physics can help convincing scientifically literate who still don't acknowledge an AGW.

\ack
This work has been partly supported by the Norwegian Meteorological Institute as a means of enlightening the general public on climate change. Part of this work is based on two posts on RealClimate.org in 2010, and I acknowledge valuable discussions in the threads of these posts.  The analysis was also based on the effort done by the European Centre for Medium-range Weather Forecasts (ECMWF) in terms of producing the ERAINT reanalysis.

\section*{References}
\bibliography{refs}

\section*{Figures and figure captions}

\begin{figure}
\includegraphics{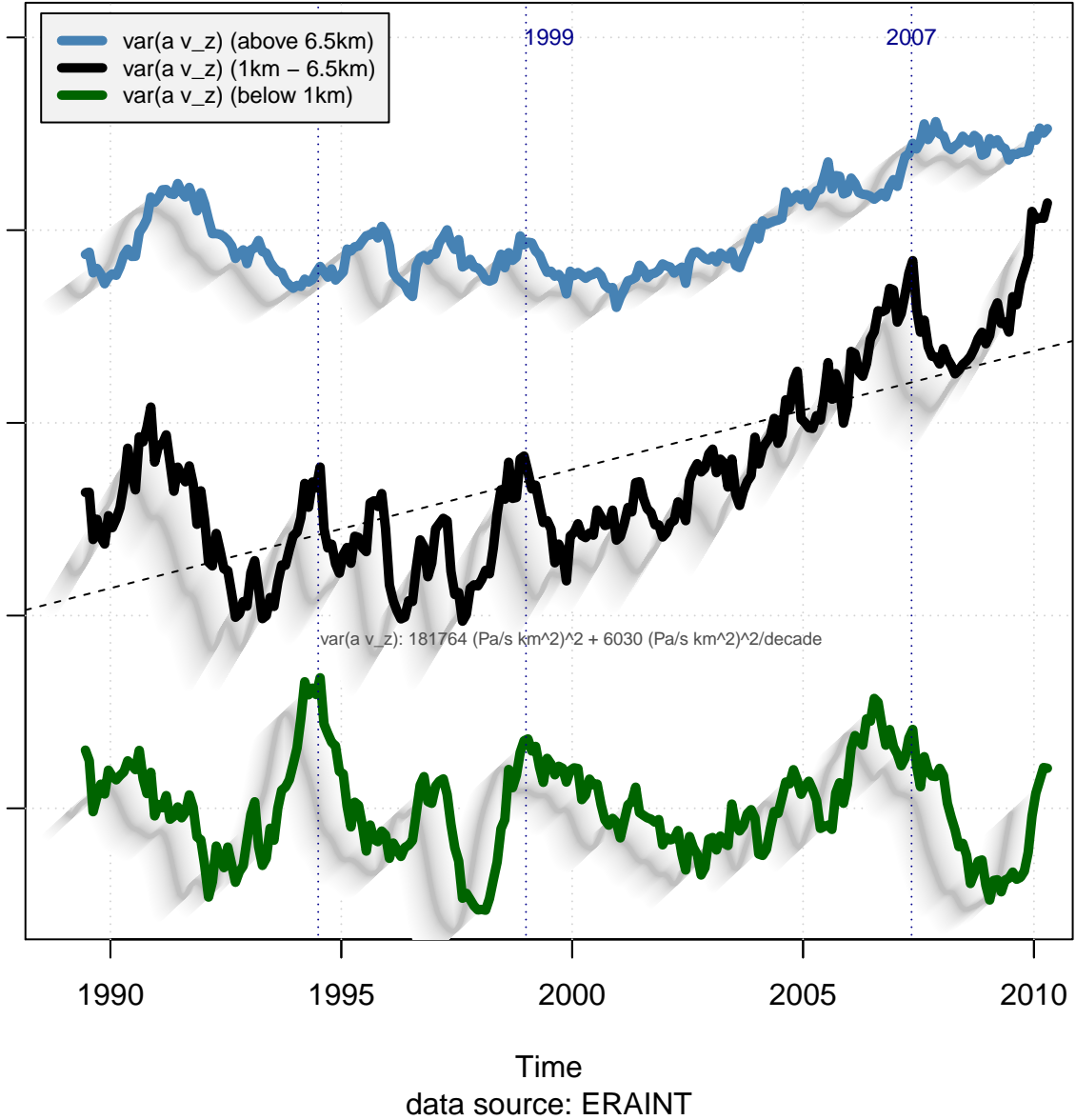}
\label{fig:1}
\caption{Curves showing the 12-month moving mean of the spatial variance of the vertical volume transport ($var(a \times v_z)$) from ERAINT.  The trend in the atmospheric overturning, $(6028 \pm 342) Pa s^{-1}/(m^2 decade)$, in the atmospheric middle levels (black) supports the notion of increased optical depth and hence and enhanced GHE. The curves are plotted here with arbitrary scales along the y-axis.}
\end{figure}

\begin{figure}
\includegraphics{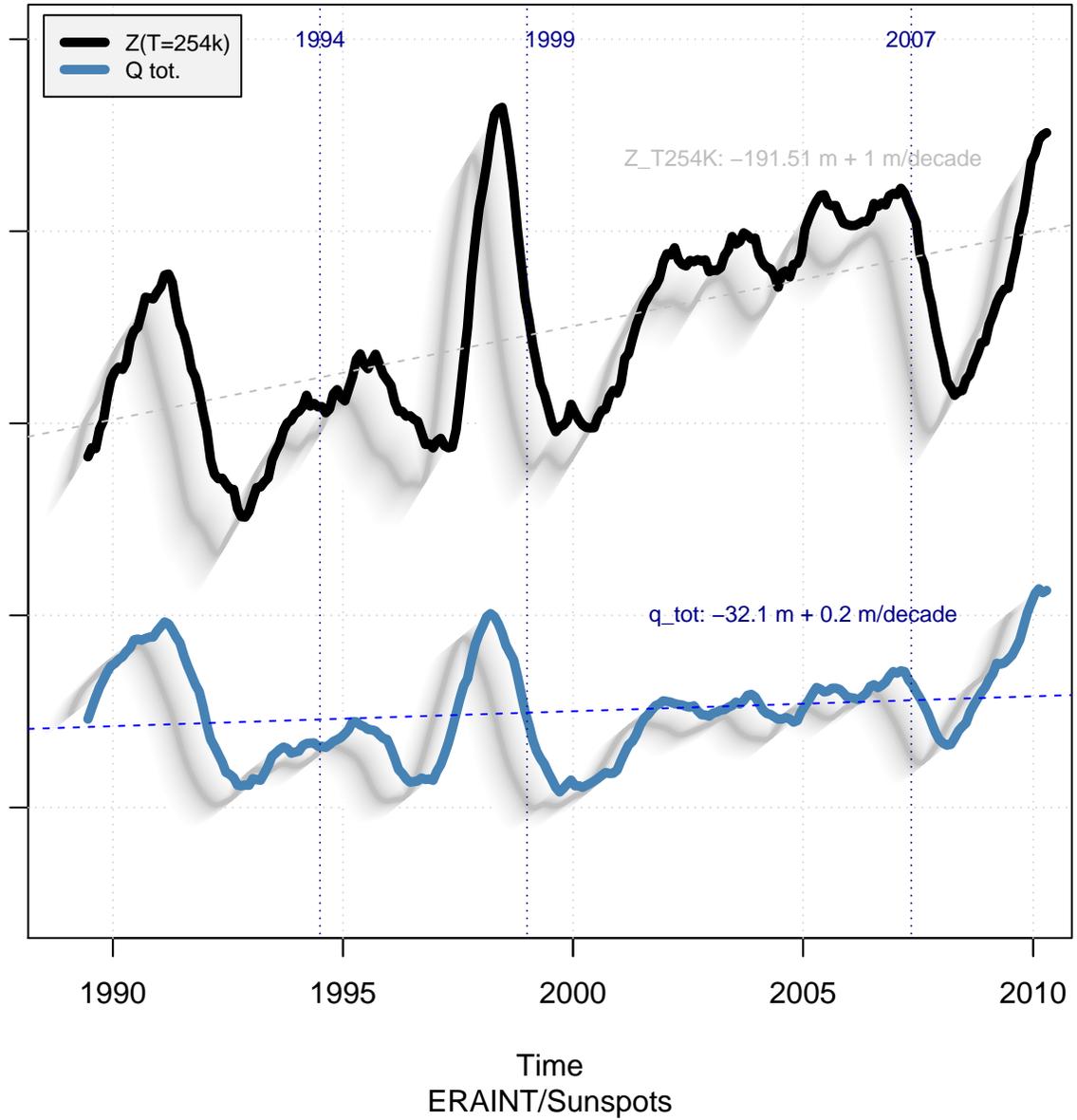}
\label{fig:2}
\caption{Curves showing the 12-month moving mean of the spatial variance of the vertical volume transport the vertical integral of total column water vapour ($q_tot$; blue), and the bulk emission level altitude ($Z_{T254K}$; black) from ERAINT. The trend in the altitude of the bulk emission level, $(0.97 \pm 0.085) m/decade$, supports the notion of increased optical depth and hence and enhanced GHE. The trend in $q_{tot}$ is $(0.015 \pm 0.005) kg/((m^2 decade)$. The curves are plotted here with arbitrary scales along the y-axis.}
\end{figure}

\clearpage

\section*{Supplementary Data}

\subsection{Supplementary analysis}

\begin{figure}
\includegraphics{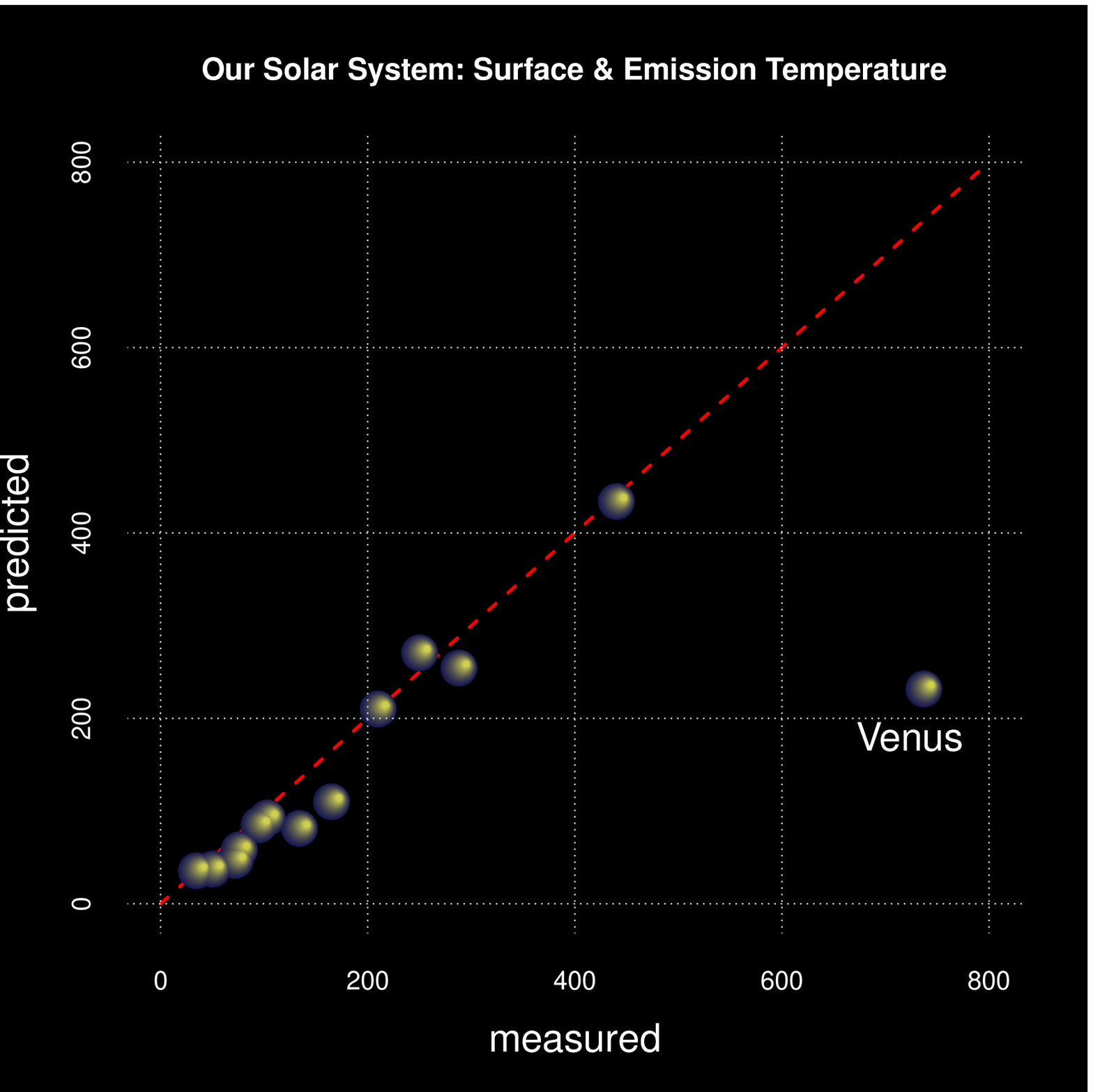}
Figure S1: Test of the energy balance model for planets and moons in our solar system. In this equation, the GHE has been neglected. Most of the predictions (y-axis) follow the observations (x-axis) and the points are close to the diagonal, except for one planet, Venus. Venus is a bright planet with a thick atmosphere mainly of CO2, and has a substantially higher surface temperature than predicted with the energy balance model. The R-script for generating this plot is given below.
\end{figure}

\begin{figure}
\includegraphics{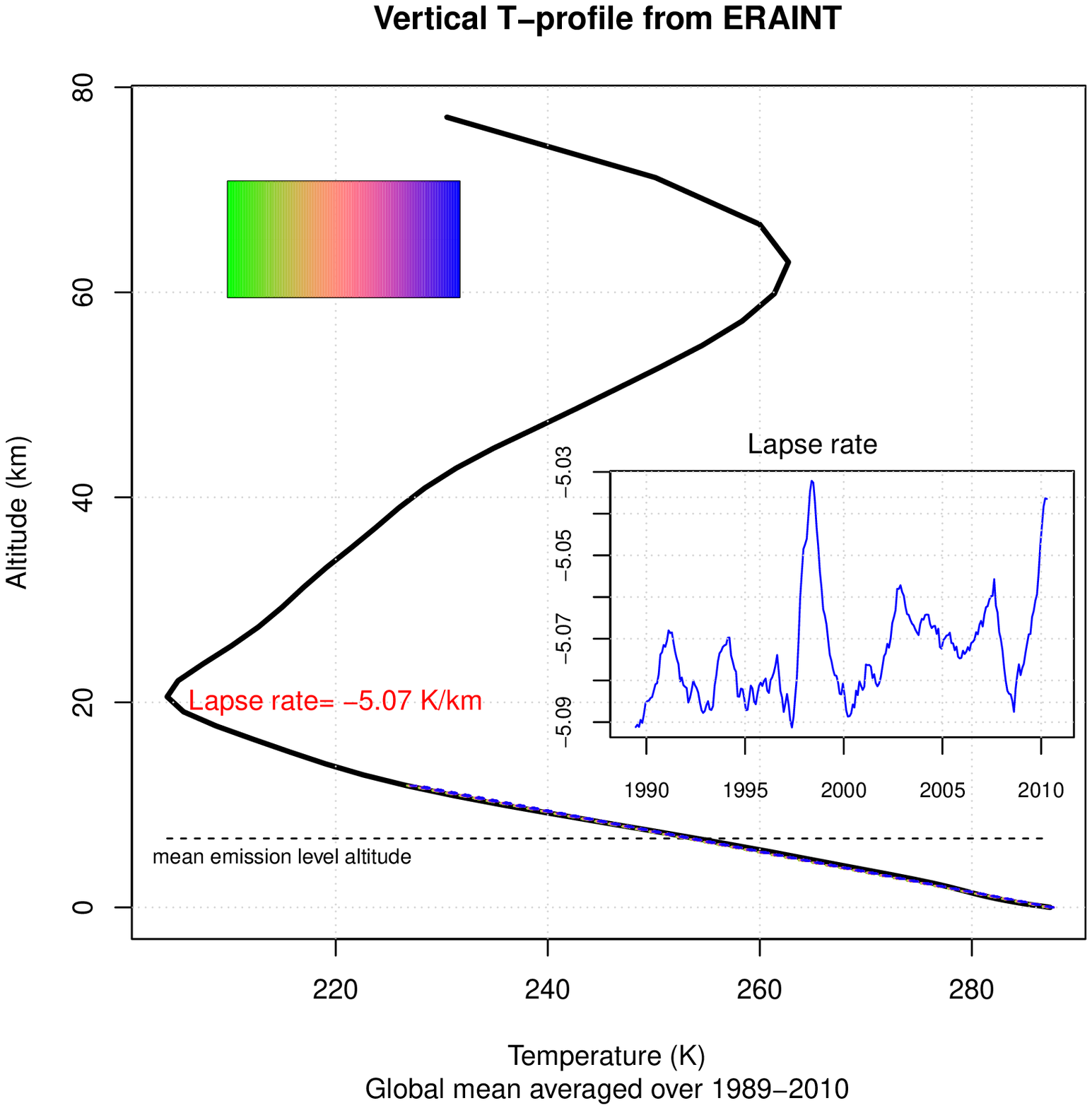}
Figure S2: The vertical temperature profile with altitude. For each ERAINT layer, the global mean was computed. The dashed horizontal line indicated the mean height for which T=254K that corresponds to the level where the bulk heat loss takes place if the energy balance equation is valid. The linear slope of the tropospheric temperature-altitude profile is estimated to have a steeper slope during the 1997-1998 El Nino, indicating the negative feedback effect associated with the lapse rate. Linear regression for the lower 12 km was used to solve for T0 and dT/dz in order to describe a linear temperature profile: T(z) = T0 + dT/dz × z. The vertical temperature profiles are shown for all years (after a 12-month moving average was used to smooth annual variations) as well as for the time mean, all of which are very similar (plotted on top of each other).  The code for generating this figure is part of the R-script used for generating Figure 1, and is listed below. 
\end{figure}

\begin{figure}
\includegraphics{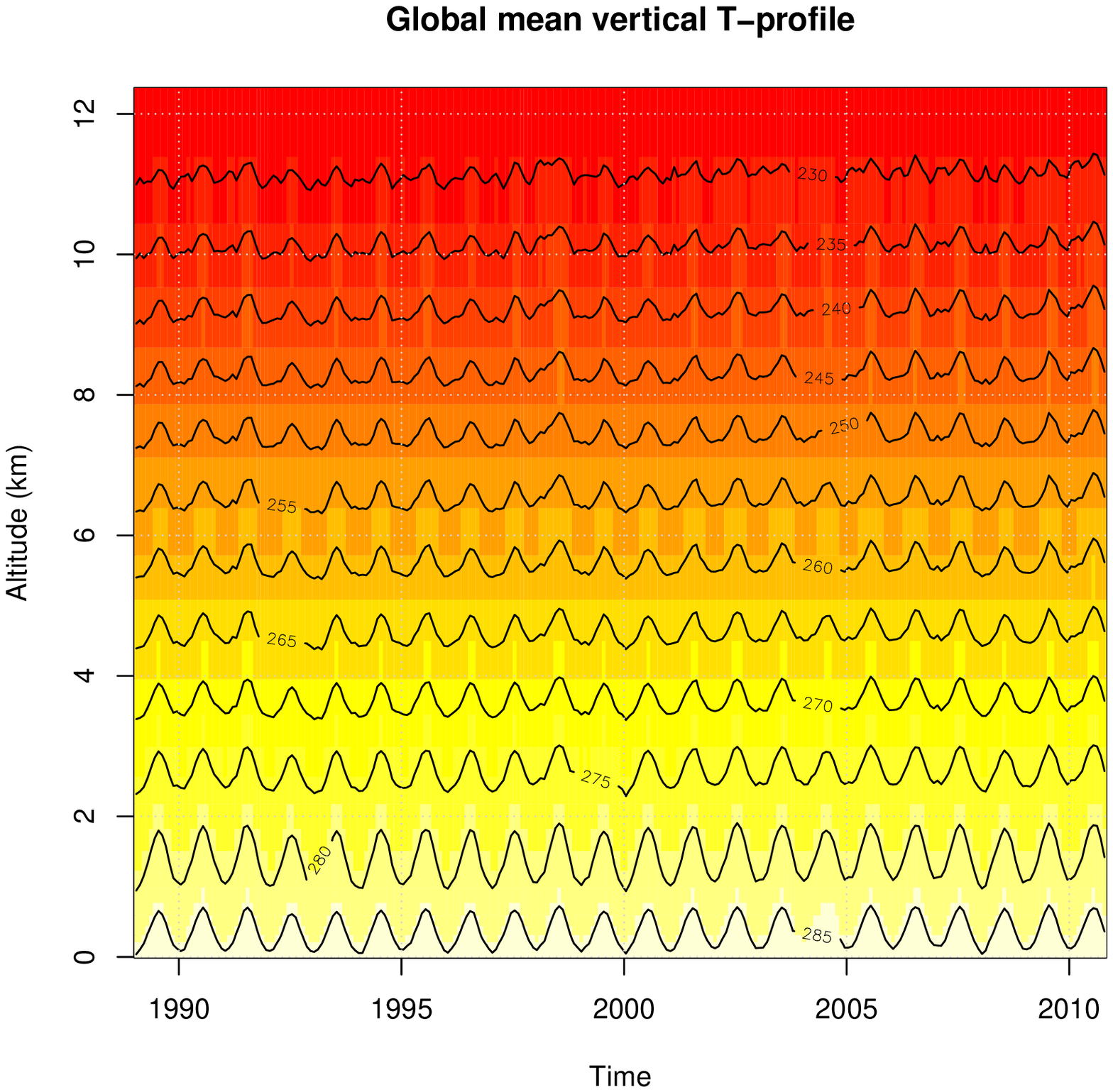}
Figure S3: The change in time of the vertical temperature profile. No clear and pronounced long-term change over time can be seen. As seen in figure S2, there are very small changes in the vertical temperature profile in the lower 12 km of the atmosphere.
\end{figure}

\begin{figure}
\includegraphics{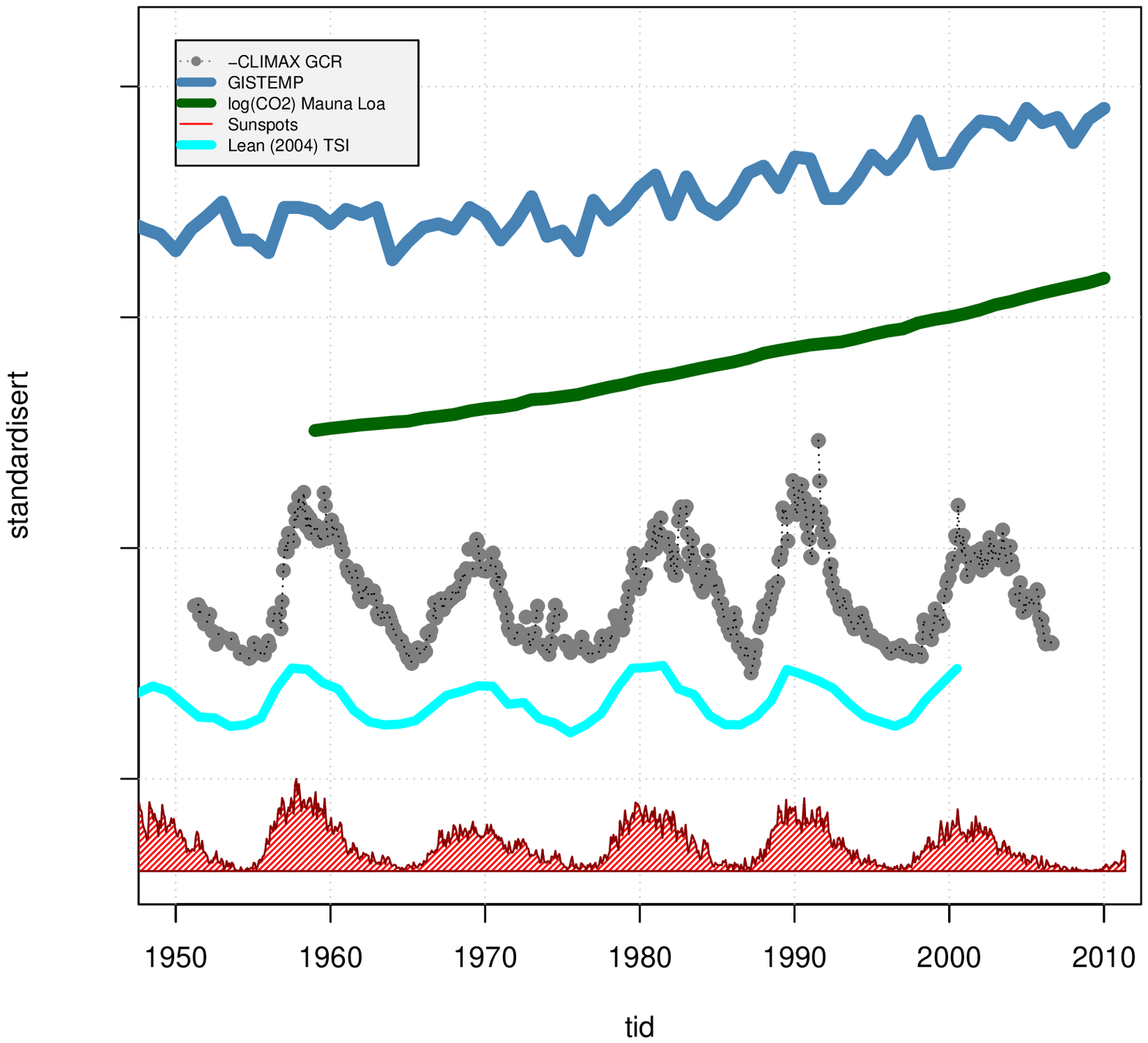}
Figure S4: Evolution of the global mean temperature (GISTEMP) compared with log(CO2) and several proxies for solar activity: Galactic cosmic rays (GCR), Lean (2004) total solar irradiance (TSI), and sunspots. A linear trend-fit to log(CO2) yields $0.041/decade \pm 0.0005$. Hence, the time before the increase is log(2)
 (0.6931472) is 16.9 decades, which if the rate of change in $Z_{T254K} = 40m/decade$, implies an altitude rise of 676m that translates to 3.4K at the ground given that dT/dz = -5K/km. This estimate does not take into account the inertia of the world oceans, and the climate sensitivity for a planet at equilibrium would suggest a greater change. Hence, the simple conceptual model predicts a climate sensitivity that is in good agreement with the range provided by the recent IPCC assessment report.
\end{figure}

\subsection{Supplementary results - IO from R window}

\begin{verbatim}
> source("erl2011.R")
[1] "T-t-prof:"
[1] "T-z-prof:"
[1] "Q:"
[1] "ERAINT model levels:"
[1] "Define atmospheric levels:"
[1] "colorbar"
> source("erl2011.R")
[1] "T-t-prof:"
[1] "T-z-prof:"
[1] "Q:"
[1] "ERAINT model levels:"
[1] "Define atmospheric levels:"
[1] "'Overturning metric':"

Call:
lm(formula = y ~ x, data = overturning.cal)

Residuals:
    Min      1Q  Median      3Q     Max 
-6273.1 -2675.7  -557.9  2216.4  8677.4 

Coefficients:
             Estimate Std. Error t value Pr(>|t|)    
(Intercept) 181764.17     206.62  879.68   <2e-16 ***
x              602.84      34.22   17.62   <2e-16 ***
---
Signif. codes:  0 ‘***’ 0.001 ‘**’ 0.01 ‘*’ 0.05 ‘.’ 0.1 ‘ ’ 1 

Residual standard error: 3273 on 249 degrees of freedom
  (11 observations deleted due to missingness)
Multiple R-squared: 0.5548,	Adjusted R-squared: 0.5531 
F-statistic: 310.4 on 1 and 249 DF,  p-value: < 2.2e-16 

[1] "'Z_T254K':"

Call:
lm(formula = Z ~ yymm)

Residuals:
     Min       1Q   Median       3Q      Max 
-1.52705 -0.66744  0.04412  0.53115  2.42788 

Coefficients:
              Estimate Std. Error t value Pr(>|t|)    
(Intercept) -1.915e+02  1.694e+01  -11.31   <2e-16 ***
yymm         9.726e-02  8.469e-03   11.48   <2e-16 ***
---
Signif. codes:  0 ‘***’ 0.001 ‘**’ 0.01 ‘*’ 0.05 ‘.’ 0.1 ‘ ’ 1 

Residual standard error: 0.8102 on 249 degrees of freedom
  (11 observations deleted due to missingness)
Multiple R-squared: 0.3463,	Adjusted R-squared: 0.3436 
F-statistic: 131.9 on 1 and 249 DF,  p-value: < 2.2e-16 

[1] "q_tot:"

Call:
lm(formula = Q ~ yymm)

Residuals:
     Min       1Q   Median       3Q      Max 
-0.83628 -0.33288 -0.03815  0.23064  1.11715 

Coefficients:
              Estimate Std. Error t value Pr(>|t|)   
(Intercept) -32.095992  10.286881  -3.120  0.00202 **
yymm          0.015549   0.005144   3.023  0.00277 **
---
Signif. codes:  0 ‘***’ 0.001 ‘**’ 0.01 ‘*’ 0.05 ‘.’ 0.1 ‘ ’ 1 

Residual standard error: 0.4921 on 249 degrees of freedom
  (11 observations deleted due to missingness)
Multiple R-squared: 0.0354,	Adjusted R-squared: 0.03153 
F-statistic: 9.138 on 1 and 249 DF,  p-value: 0.002765 

[1] "colorbar"
\end{verbatim}

\subsection{Data}

The ERAINT data was extracted from the ECMWF MARS archive as a netCDF file with 1º× 1º horizontal resolution, monthly mean values for the interval January 1989 to October 2010. The unit of the vertical velocity was Pa/s, taken at the models vertical levels: 250, 300, 350, 400, 450, 500, 550, 600, 650, 700, 750, 775, 800, 825, 850, 875, 900, 925, 950, 975, 1000 hPa. For temperature, all 60 model levels were used. 

\subsection{Methods \& Scripts}

\subsubsection{Data extraction in Ferret}

The application ferret (http://ferret.pmel.noaa.gov/Ferret/), from the U.S. National Oceanic and Atmospheric Administration's (NOAA) Pacific Marine Environment Lab (PMEL) was  used for making somee of the plots as well as for post-processing and extracting some of the reanalysis. The ferret scripts are provided below:

Ferret script for Ferret script for plotting vertical temperature profiles and extracting temperature profiles and atmospheric moisture:

\begin{verbatim}
use ERAINT_monthly-T-z.nc
set reg/z=0:10 !t="01-Jan-2000":"31-Dec-2010"

let temp = TAIR[x=@ave,y=@ave]
! extract global mean temperature for each horizontal level
save/file=horizmeanT.nc temp

! find the emission level height:
let lev= temp[z=@loc:255K]

! extract the global mean vertical temperature structure:
let prof= temp[l=@ave]

fill/vlim=0:7:0.1 temp[l=@sbx:24]
plot/over/nolab lev[l=@sbx:24]

!Save extracted data:
save/file=Tprofile.nc prof
save/file=emissionlevel.nc lev

use eraint_q-verticalint.nc
plot q[x=@ave,y=@ave,l=@sbx:12]
\end{verbatim}

\subsubsection{Open source code and scripts for more advanced analysis and plots}

The R-environment (http://cran.r-project.org) was used for generating Figures 1 \& 2, and the R script for generating this plot is provided below:

\begin{verbatim}
# R.E. Benestad, 12.01.2011.

library(clim.pact)  # Available from http://cran.r-project.org

# Simple formula for estimating the pressure at a given height 
p.hydrostatic <- function (h,p0 = 1013.250,Temp = 288,g = 9.81,
                           k = 1.38e-23,M = 0.027/6.022e+23) {
    p <- p0 * exp(-(M * g * h)/(k * Temp))
    p
}


stand <- function (x, m = NULL, s = NULL) {
    if (is.null(m)) 
        m <- mean(x, na.rm = TRUE)
    if (is.null(s)) 
        s <- sd(x, na.rm = TRUE)
    labels <- seq(min(10 * round(floor(x)/10, 1), na.rm = TRUE), 
        max(10 * round(ceiling(x)/10, 1), na.rm = TRUE), length = 20)
    x <- (x - m)/s
    at <- (round(labels, 2) - m)/s
    attr(x, "m") <- m
    attr(x, "s") <- s
    attr(x, "axis.at") <- at
    attr(x, "axis.labels") <- round(labels, 2)
    x
}

ma.filt <- function (x, n) {
    y <- filter(x, rep(1, n)/n)
    y
}

sunspots <- function(url="http://sidc.oma.be/DATA/monthssn.dat") {
  asciidata <- readLines(url)
  for (i in 1:6) asciidata[i] <- paste(asciidata[i],"NA")
  for (i in (length(asciidata)-20):length(asciidata)) {
    asterisk <- instring("*",asciidata[i])
    #print(paste("i=",i,"astersk=",asterisk,asciidata[i]))
    if (length(asterisk)>0) {
      if (asterisk[1]>0) {
        if (length(asterisk)==1) asciidata[i] <-
          substr(asciidata[i],1,asterisk[1]-1) else
        if (length(asterisk)==2) asciidata[i] <-
          paste(substr(asciidata[i],1,asterisk[1]-1),
                substr(asciidata[i],asterisk[1]+1,asterisk[2]-1))
      }
    } 
    if (i >= length(asciidata)-5) asciidata[i] <- paste(asciidata[i],"NA")
  }
  writeLines(asciidata,"SIDC-sunspotnumber.txt")
  Rs <- read.table("SIDC-sunspotnumber.txt",header=FALSE,
                   col.names=c("yyyymm","year","sunspotnumber","smoothed"))
  attr(Rs,"description") <-
    list(src="monthly sunspot number from ROYAL OBSERVATORY OF BELGIUM",url=url)
  Rs
}

colorbar <-  function(yymm,fig=c(0.33,0.66,0.30,0.34),breaks,col) {
  print("colorbar")
  fig.old <- c(0,1,0,1)
  par(fig=fig,new=TRUE,mar=rep(0,4),xaxt="n",yaxt="n",cex.axis=0.5)
  colbar <- cbind(c(breaks[-1]),c(breaks[-1])); d <- dim(colbar)
  #print(d); print(length(yymm))
  image(yymm,c(0,1),colbar,breaks=breaks,col=col)
  par(fig=fig.old,new=TRUE)
}


prettygraph <- function(t,y,col="black",tshadeoffset=0.1,yshadeoffset=0.1) {
  greys <- rgb( (seq(1,0.06,length=100)^0.1),
                (seq(1,0.06,length=100)^0.1),
                (seq(1,0.06,length=100)^0.1) )
  toffs <- tshadeoffset*(max(t,na.rm=TRUE)- min(t,na.rm=TRUE))*0.25
  yoffs <- yshadeoffset*(max(y,na.rm=TRUE)- min(y,na.rm=TRUE))
  for (i in 1:100) {
    ii <- ( (i - 100)/100 )
    lines(t-toffs+ii*toffs,gauss.filt(y-yoffs+ii*yoffs,5),lwd=2,col=greys[i])
    lines(t-toffs-ii*toffs,gauss.filt(y-yoffs-ii*yoffs,5),lwd=2,col=greys[i])
  }
  lines(t,y,lwd=5,col=col)
}



# The netCDF-file with vertical velocity from ERAINT was taken
# from ECMWF's MARS archive.

tempfil <- "~/ferret_scripts/horizmeanT.nc"
proffil <- "~/ferret_scripts/Tprofile.nc"
qfil <- "~/ferret_scripts/humidity.nc"

if (!file.exists("atmOverturning.Rdata")) {
  ncid <- open.ncdf("~/data/ERAINT/eraint_w.nc")
  lon = get.var.ncdf(ncid,"longitude")
  lat=get.var.ncdf(ncid,"latitude")
  lev=get.var.ncdf(ncid,"levelist")
  tim=get.var.ncdf(ncid,"time")

  nz <- length(lev); nt <- length(tim); nx <- length(lon); ny <- length(lat)
  wa <- matrix(rep(NA,nz*nt),nz,nt)

  # Area of the grid boxes
  box.area <- rep(cos(pi*lat/180)* pi*min(diff(lon))/180*
                  abs(pi*min(diff(lat)))/180*(6.378e03)^2,nx)
  dim(box.area) <- c(ny,nx); box.area <- t(box.area)
  
  for (iz in 1:nz){
    w = get.var.ncdf(ncid,"w",start=c(1,1,iz,1),count=c(nx,ny,1,nt))

    ysrt <- order(lat)
    image(lon,lat[ysrt],box.area[,ysrt],
          main=paste("Check: level=",lev[iz],"hPa"))
    addland()
    contour(lon,lat[ysrt],w[,ysrt,1],add=TRUE)
  
    for (it in  1:nt) {
      X <- c(w[,,it]*box.area)
      wa[iz,it] <- var(X)
    }
  }
  close.ncdf(ncid)
  save(file="atmOverturning.Rdata",wa,tim,lev,lon,lat)
} else {
  load ("atmOverturning.Rdata")
  nz <- length(lev); nt <- length(tim); nx <- length(lon); ny <- length(lat)
}

Z <- seq(-10,100000,by=10)
heights <- round(0.1*approx(x=p.hydrostatic(Z),y=Z,xout=lev)$y)*10

upper <- heights >= 6500
middle <- (heights < 6500) & heights >= 1000
lower <- heights < 1000
date <- caldat( tim/24 + julday(1,1,1900) )
yymm <- date$year + (date$month - 0.5)/12

print("T-t-prof:")
ncid2 <- open.ncdf(tempfil)
temp2 <- get.var.ncdf(ncid2,"TEMP")
tim2 <-  get.var.ncdf(ncid2,"TIME")
lev <-  get.var.ncdf(ncid2,"LEVELIST")
close.ncdf(ncid2)

date2 <- caldat( tim2/24 + julday(1,1,1900) )
yymm2 <- date2$year + (date2$month - 0.5)/12

print("T-z-prof:")
ncid3 <- open.ncdf(proffil)
temp3 <- get.var.ncdf(ncid3,"PROF")
close.ncdf(ncid3)
nt2 <- length(tim2)

print("Q:")
ncid4 <- open.ncdf(qfil)
Q <- get.var.ncdf(ncid4,"Q")
close.ncdf(ncid4)

print("ERAINT model levels:")
# http://www.ecmwf.int/products/data/technical/model_levels/model_def_60.html
levpres <- read.table("eraint-model.levels.txt",header=TRUE)
alt3 <- round(0.1*approx(x=p.hydrostatic(Z),y=Z,xout=levpres$ph)$y)*10
t254k <- rep(NA,nt2)

print("Define atmospheric levels:")
troposphere <- alt3 < 12000
alt <- alt3[troposphere]
temp2 <- temp2[troposphere,]

overturning <- ma.filt(colMeans(wa[middle,]) - mean(wa[middle,]),12)
overturning.cal <- data.frame(y=ma.filt(colMeans(wa[middle,]),12),
                              x=yymm - mean(yymm))


# Graphics
# ERL Figure 1:

par(col.axis="white")
y <- stand(overturning) + 1.5; 
plot(yymm, y,type="n",
     main="Atmospheric 'overturning' anomaly",
     ylab="",xlab="Time",xlim <- c(1989,2010),ylim=c(-3,6),
     sub="data source: ERAINT")
par(col.axis="black")
axis(1)
grid()

prettygraph(yymm, y)
prettygraph(yymm, 0.5*stand(ma.filt(colMeans(wa[upper,]),12))+4,col="steelblue")
prettygraph(yymm, 0.5*stand(ma.filt(colMeans(wa[lower,]),12))-2,col="darkgreen")

abline(lm(y ~yymm),lty=2)
Avz.trend <- summary(lm(y ~ x,data=overturning.cal))$coefficients
c <- round(Avz.trend)
text(2000,-0.25,paste("var(a v_z):",c[1],"(Pa/s km^2)^2 +",10*c[2],"(Pa/s km^2)^2/decade"),
     cex=0.6,col="grey30")
print("'Overturning metric':")
print(summary(lm(y ~ x,data=overturning.cal)))

lines(rep(1994.5,2),c(-10000,15000),lwd=1,lty=3,col="darkblue")
lines(rep(1999,2),c(-10000,15000),lwd=1,lty=3,col="darkblue")
lines(rep(2007.35,2),c(-10000,15000),lwd=1,lty=3,col="darkblue")
text(1994.5,6,"1994",cex=0.75,col="darkblue")
text(1999.5,6,"1999",cex=0.75,col="darkblue")
text(2007.35,6,"2007",cex=0.75,col="darkblue")

legend(1988.5,6.25,
       c("var(a v_z) (above 6.5km)","var(a v_z) (1km - 6.5km)",
         "var(a v_z) (below 1km)"),
       col=c("steelblue","black","darkgreen"),
       lwd=5,bg="grey95",cex=0.75)

dev2bitmap("erl2011-fig1.png",res=150)
dev2bitmap("erl2011-fig1.jpg",type="jpeg",res=200)
dev.copy2eps(file="erl2011-fig1.eps")
dev2bitmap("erl2011-fig1.pdf",type="pdfwrite")

x11()
# ERL Figure 2:

par(col.axis="white")

# Emission level height:
for (i in 1:nt) t254k[i] <- approx(x=temp2[,i],y=alt,xout=254)$y
Z <- stand(ma.filt(t254k,12))+3;

# Total water vapour content:
Q <- 0.5*stand(ma.filt(Q,12)) - 1

plot(yymm, y,type="n",lwd=4,
     main="Atmospheric bulk emission level and moisture",
     ylab="",xlab="Time",xlim <- c(1989,2010),ylim=c(-3,6),
     sub="ERAINT/Sunspots")
par(col.axis="black")
axis(1)
grid()

prettygraph(yymm, Z)
prettygraph(yymm, Q,col="steelblue")


# Emission level height trend:
abline(lm(Z ~ yymm),lty=2,col="grey")
t254k.trend <- summary(lm(Z ~ yymm))
c <- round(t254k.trend$coefficients,2)
text(2005,4.75,paste("Z_T254K:",c[1],"m +",10*c[2],"m/decade"),
     cex=0.75,col="grey")
print("'Z_T254K':")
print(summary(lm(Z ~ yymm)))

# Total water vapour content trend:
abline(lm(Q ~ yymm),lty=2,col="blue")
t254k.trend <- summary(lm(Q ~ yymm))
c <- round(t254k.trend$coefficients,2)
text(2005,0,paste("q_tot:",c[1],"m +",10*c[2],"m/decade"),
     cex=0.75,col="darkblue")
print("q_tot:")
print(summary(lm(Q ~ yymm)))

lines(rep(1994.5,2),c(-10000,15000),lwd=1,lty=3,col="darkblue")
lines(rep(1999,2),c(-10000,15000),lwd=1,lty=3,col="darkblue")
lines(rep(2007.35,2),c(-10000,15000),lwd=1,lty=3,col="darkblue")
text(1994.5,6,"1994",cex=0.75,col="darkblue")
text(1999.5,6,"1999",cex=0.75,col="darkblue")
text(2007.35,6,"2007",cex=0.75,col="darkblue")

legend(1988.5,6.25,
       c("Z(T=254k)","Q tot."),
       col=c("black","steelblue"),
       lty=c(1,1),lwd=5,bg="grey95",cex=0.75)

dev2bitmap("erl2011-fig2.png",res=150)
dev2bitmap("erl2011-fig2.jpg",type="jpeg",res=200)
dev.copy2eps(file="erl2011-fig2.eps")
dev2bitmap("erl2011-fig2.pdf",type="pdfwrite")


# Supporting diagnostics

cols <- rgb(sin(pi*seq(0,1,length=nt2)),seq(1,0,length=nt2),seq(0,1,length=nt2))

x11()
plot(temp3,alt3/1000,type="l",lwd=3,
     main="Vertical T-profile from ERAINT",
     sub="Global mean averaged over 1989-2010",
     xlab="Temperature (K)",ylab="Altitude (km)")
grid()
lines(range(temp3),rep(mean(t254k)/1000,2),lty=2)
text(215,5,"mean emission level altitude",cex=0.75)

trop <- (alt3/1000) < 12
t <- temp3[trop]; z <- alt3[trop]/1000; srtz <- order(z)
calibr <- data.frame(t=t[srtz],z=z[srtz])
lapserate <- lm(t ~ z,data=calibr)
lines(predict(lapserate,newdata=calibr),z[srtz],lty=2,col="red")
text(220,20,paste("Lapse rate=",
                  round(summary(lapserate)$coefficients[2],2),"K/km"),col="red")

LR <- rep(NA,nt2)
temp4 <- temp2; d <- dim(temp4)
z <- alt/1000; srtz <- order(z)
for (j in 1:d[1]) temp4[j,] <- ma.filt(temp4[j,],12)
for (i in 1:nt2) {
  t <- temp4[,i]
  if (sum(is.finite(t))==d[1]) {
    calibr <- data.frame(t=t[srtz],z=z[srtz])
    lapserate <- lm(t ~ z,data=calibr)
    lines(predict(lapserate,newdata=calibr),z[srtz],lty=2,col=cols[i])
    LR[i] <- summary(lapserate)$coefficients[2]
  }
}

text(265,45,"Lapse rate")

colorbar(yymm=yymm2,fig=c(0.2,0.4,0.70,0.80),breaks=0:nt2,col=cols)

fig.old <- c(0,1,0,1)
par(fig=c(0.53,0.93,0.32,0.55),new=TRUE,cex.axis=0.75,xaxt="s",yaxt="s",
    cex.main=0.75)
plot(yymm2,LR,type="l",xlab="",ylab="",col="blue")
grid()
par(fig=fig.old,new=TRUE)

dev2bitmap("erl2011-figS2.png",res=150)
dev2bitmap("erl2011-figS2.jpg",type="jpeg",res=200)
dev.copy2eps(file="erl2011-figS2.eps")
dev2bitmap("erl2011-figS2.pdf",type="pdfwrite")


x11()
plot(temp3,alt3/1000,type="l",lwd=3,ylim=c(0,12),
     main="Vertical Temperature Profile",
     sub="Global mean averaged over 1989-2010 (ERAINT)",
     xlab="Temperature (K)",ylab="Altitude (km)")
grid()
lines(range(temp3),rep(mean(t254k)/1000,2),lty=2)
text(215,7,"mean emission level altitude",cex=0.75)

trop <- (alt3/1000) < 12
t <- temp3[trop]; z <- alt3[trop]/1000; srtz <- order(z)
calibr <- data.frame(t=t[srtz],z=z[srtz])
lapserate <- lm(t ~ z,data=calibr)
lines(predict(lapserate,newdata=calibr),z[srtz],lty=2,col="red")
text(220,20,paste("Lapse rate=",
                  round(summary(lapserate)$coefficients[2],2),"K/km"),col="red")

LR <- rep(NA,nt2)
temp4 <- temp2; d <- dim(temp4)
z <- alt/1000; srtz <- order(z)
for (j in 1:d[1]) temp4[j,] <- ma.filt(temp4[j,],12)
for (i in 1:nt2) {
  t <- temp4[,i]
  if (sum(is.finite(t))==d[1]) {
    calibr <- data.frame(t=t[srtz],z=z[srtz])
    lapserate <- lm(t ~ z,data=calibr)
    lines(predict(lapserate,newdata=calibr),z[srtz],lty=2,col=cols[i])
    LR[i] <- summary(lapserate)$coefficients[2]
  }
}

text(265,45,"Lapse rate")

x11()
srtz <- order(alt)
image(yymm2,alt[srtz]/1000,t(temp2[srtz,]),
      main="Global mean vertical T-profile",
      ylab="Altitude (km)",xlab="Time")
contour(yymm2,alt[srtz]/1000,t(temp2[srtz,]),add=TRUE)
grid()

dev2bitmap("erl2011-figS3.png",res=150)
dev2bitmap("erl2011-figS3.jpg",type="jpeg",res=200)
dev.copy2eps(file="erl2011-figS3.eps")
dev2bitmap("erl2011-figS3.pdf",type="pdfwrite")
\end{verbatim}

The R script for generating Figure S1 is provided below:

\begin{verbatim}
largesymbols <- function(x,y,col="blue",cex=3) {
  N <- 100
  others <- seq(0.3,0.9,length=N)^2 
  if (col=="blue") cols <- rgb(others,others,sqrt(seq(0.1,1,by=N))) else
  if (col=="red") cols <- rgb(sqrt(seq(0.1,1,by=N)),others,others) else
  if (col=="green") cols <- rgb(others,sqrt(seq(0.1,1,by=N)),others) else
  if (col=="yellow") cols <- rgb(sqrt(seq(0.1,1,by=N)),sqrt(seq(0.1,1,by=N)),others)
  for (i in 1:N) {
    xofs <- 0.01*i/N*max(x,na.rm=TRUE); yofs <- 0.01*i/N*max(y,na.rm=TRUE)
    sizefact <- 1 - (i-1)/N*0.9
    points(x+xofs,y+yofs,pch=19,cex=cex*sizefact,col=cols[i])
  }
}

T.e <- function(R,A,S0=1367,rho=5.67e-8) {
  S <- S0/(R^2)
  T.e <- (S*(1-A)/4/rho)^0.25
  T.e
}



np <- 13
pch <- rep(19,np)
col <- rep("grey20",np)
col[3] <-"blue"
pch[4] <- 21
pch[11:13] <- 4 

X <- matrix(rep(NA,np*3),np,3)
rownames(X)  <- c("Mercury","Venus","Earth","Moon","Mars","Jupiter",
                  "Saturn","Uranus","Neptun","Pluto",
                  "Titan","Europa","Triton")
colnames(X) <- c("R","Albedo","Mean.T")

# From:
#Planetary database
#Source for planetary data, and some of the data on
#the moons, is http://nssdc.gsfc.nasa.gov/planetary/factsheet/
# ~/data/planets.txt
X[1,] <- c(0.387, 0.119, 440.0)
X[2,] <- c(0.723, 0.750, 737.0)
X[3,] <- c(1.000, 0.306, 288.0)
X[4,] <- c(1.000, 0.110, 0.5*(100+400))
X[5,] <- c(1.524, 0.250, 210.0)
X[6,] <- c(5.203, 0.343, 165.0)
X[7,] <- c(9.539, 0.342, 134.0)
X[8,] <- c(19.181,0.300,  76.0)
X[9,] <- c(30.058,0.290,  72.0)
X[10,] <- c(39.5, 0.500,  50.0)
X[11,] <- c(9.539,0.21,   95.0)
X[12,] <- c(5.203,0.67,  103.0)
X[13,] <- c(30.058,0.76,  34.5)

# From http://www.astro-tom.com/getting_started/planet_classification.htm
#      http://www.solarviews.com/eng/moon.htm
# A from Houghton (1986) The Physics of Atmospheres
# http://www.astronomytoday.com/astronomy/mercury.html
# http://www.universetoday.com/guide-to-space/the-moon/moon-albedo/
# 
#R <-   c(0.387,0.723, 1,   1,    1.524, 5.203, 9.539, 19.181, 30.058, 39.5)
#A <-   c(0.06, 0.77,  0.30,0.12,0.15, 0.58, )
#Tsd <- c(350,  480,   15, 107, -23,   -150,  -180,  -214,   -220,   -230)
#Tsn <- c(-170, 480,   15, -153,  -23,   -150,  -180,  -214,   -220,   -230)
#Ts <-  0.5*(Tsd + Tsn) + 273.15

srt <- order(X[,1])
R <- X[srt,1]; A <- X[srt,2]; Ts <- X[srt,3];
col <- col[srt]; pch <- pch[srt]

x11()
par(ps=12,bg="black",col.axis="white",col.main="white",col.lab="white",
    cex.lab=1.5)
plot(Ts,T.e(R,A),pch=pch,col=col,cex=1.5,
     main="Our Solar System: Surface & Emission Temperature",
     sub="GHE affects observed temperature, but is not accounted for in predictions",
     xlab="measured",ylab="predicted",xlim=c(0,800),ylim=c(0,800))
lines(c(0,800),c(0,800),col="red",lty=2,lwd=2)
largesymbols(Ts,T.e(R,A),col="blue",cex=3)
grid()
text(725,180,"Venus",cex=1.5,col="white")
\end{verbatim}

The R script for generating Figure S4 is provided below:

\begin{verbatim}
adj <- function(y1,y2,x1,x2) {
  ii <- (x1 >= min(x2,na.rm=TRUE)) & (x1 <= max(x2,na.rm=TRUE))
  rng.y1 <- range(y1[ii],na.rm=TRUE)
  rng.y2 <- range(y2,na.rm=TRUE)

  y1 <- (y1 - rng.y1[1])/(rng.y1[2] - rng.y1[1]) * (rng.y2[2] - rng.y2[1]) + rng.y2[1]
  y1
}



  print("Get GISS from URL")
  test.read <- readLines("http://data.giss.nasa.gov/gistemp/graphs/Fig.A2.txt")
  if (now("c(day,mon,year)")[2]==1) iglines <- 5 else
                                    iglines <- 6
  nrows <- length(test.read) - iglines

  writeLines(test.read, "GISS-T2m.mon.txt")


t2m <- read.table("GISS-T2m.mon.txt", skip = 4, col.names = c("Year",
        "Annual.Mean", "X5.year.mean"), nrows = nrows, na.strings="*")
t2m$X5.year.mean <- as.character(t2m$X5.year.mean)
t2m$X5.year.mean[is.element(t2m$X5.year.mean, "*")] <- NA
t2m$X5.year.mean <- as.numeric(t2m$X5.year.mean)
t2m$Annual.Mean <- t2m$Annual.Mean -
  mean(t2m$Annual.Mean[is.element(t2m$Year,1961:1990)])

R <- read.table("~/data/indices/sunspot_num.dat",
                col.names=c("year","month","number"))
R$number[R$number<0] <- NA
yymm <- R$year+(R$month-0.5)/12


#rm(list=ls())
#library(chron)

url1 <- "ftp://ulysses.sr.unh.edu/NeutronMonitor/DailyAverages.1951-.txt"
url2 <- "ftp://ftp.ngdc.noaa.gov/STP/GEOMAGNETIC_DATA/AASTAR/aaindex"

print(paste("Reading GCR data from",url1,"..."))
a <- readLines(url1)
print("Saving GCR data in local file 'neutron2.dat'; skipping the header:")
#print(a[1:43])
a <- a[44:length(a)]
writeLines(a,"neutron2.dat")

gcr <- read.table("neutron2.dat",header=T, as.is=T,
                  col.names=c("Date","SecsOf1904","Climax","Huancayo.NoGC",
                             "Huancayo.GC","Haleakala.IGY","Haleakala.S/M"))
climax <- filter(as.numeric(gcr$Climax),rep(1,30)/30)
nt.gcr <- length(climax)
climax <- climax[seq(15,nt.gcr,by=30)]
gcr.yy <- as.numeric(substr(as.character(gcr$Date),7,10))
gcr.mm <- as.numeric(substr(as.character(gcr$Date),1,2))
gcr.dd <- as.numeric(substr(as.character(gcr$Date),4,5))
jday.gcr  <- gcr.yy + (gcr.mm-1)/12 + (gcr.dd-0.5)/365.25
jday.gcr <- jday.gcr[seq(15,nt.gcr,by=30)]


url <- 'http://gcmd.nasa.gov/records/GCMD_NOAA_NCDC_PALEO_2004-035.html'
ftp <- 'ftp://ftp.ncdc.noaa.gov/pub/data/paleo/climate_forcing/solar_variability/lean2000_irradiance.txt'


if (!file.exists("Lean2004.txt")) {
  print("Get Lean (2004) from URL")
  Lean2004 <- readLines(ftp)
  first <- grep("YEAR",Lean2004)
  a <- Lean2004[first:length(Lean2004)]
  meta.data <- Lean2004[1:(first-1)]              
  writeLines(a,"Lean2004.txt")
}
Lean2004 <- read.table("Lean2004.txt",header=TRUE)

# Re

par(col.axis="white")
plot(c(1950,2010),c(-2,16),type="n",xlab="tid",ylab="standardisert")
par(col.axis="black")
axis(1)
grid()


polygon(c(yymm,yymm[length(yymm)],yymm[1]),c(2*R$number/max(R$number,na.rm=TRUE),0,0)-2,
        col="red",border="darkred",density=40)

lines(t2m$Year,stand(t2m$Annual.Mean)+12,lwd=7,col="steelblue")
points(jday.gcr,stand(-climax)+4,pch=19,col="grey50")
lines(jday.gcr,stand(-climax)+4,lty=3)

co2.test <- readLines("http://cdiac.ornl.gov/ftp/trends/co2/maunaloa.co2")
nrows.co2 <- length(co2.test) - 20
Mauna.Loa <- read.table("http://cdiac.ornl.gov/ftp/trends/co2/maunaloa.co2",
                        skip=16,nrows=nrows.co2)
Mauna.Loa[Mauna.Loa<0] <- NA

co2 <- stand(log(Mauna.Loa[,14]))
co2.trend <- data.frame(y=log(Mauna.Loa[,14]),t=Mauna.Loa[,1] - mean(Mauna.Loa[,1]))
lines(Mauna.Loa[,1],co2+9,type="l",lwd=7,col="darkgreen")
print(summary(lm(y ~ t, data=co2.trend)))

lines(Lean2004$YEAR,stand(Lean2004$"X11yrCYCLE.BKGRND"),lwd=5,col="cyan") 

legend(1950,16,c("-CLIMAX GCR      ","GISTEMP      ","log(CO2) Mauna Loa     ",
                 "Sunspots     ", "Lean (2004) TSI    "),
       pch=c(19,26,26,26,26),col=c("grey50","steelblue","darkgreen","red","cyan"),lwd=c(1,5,5,1,5),
       lty=c(3,1,1,1,1),cex=0.6,bg="grey95")
\end{verbatim}

\end{document}